# Morphological Neuron Classification Using Machine Learning

Xavier Vasques[1,2]*, Laurent Vanel[2], Guillaume Villette[2] and Laura Cif[3,4]

[1] Laboratoire de Recherche en Neurosciences Cliniques, Saint-André-de-Sangonis, France, [2] International Business Machines Corporation Systems, Paris, France, [3] Département de Neurochirurgie, Hôpital Gui de Chauliac, Centre Hospitalier Régional Universitaire de Montpellier, Montpellier, France, [4] Université de Montpellier 1, Montpellier, France

Classification and quantitative characterization of neuronal morphologies from histological neuronal reconstruction is challenging since it is still unclear how to delineate a neuronal cell class and which are the best features to define them by. The morphological neuron characterization represents a primary source to address anatomical comparisons, morphometric analysis of cells, or brain modeling. The objectives of this paper are (i) to develop and integrate a pipeline that goes from morphological feature extraction to classification and (ii) to assess and compare the accuracy of machine learning algorithms to classify neuron morphologies. The algorithms were trained on 430 digitally reconstructed neurons subjectively classified into layers and/or m-types using young and/or adult development state population of the somatosensory cortex in rats. For supervised algorithms, linear discriminant analysis provided better classification results in comparison with others. For unsupervised algorithms, the affinity propagation and the Ward algorithms provided slightly better results.

## INTRODUCTION

The quantitative characterization of neuronal morphologies from histological digital neuronal reconstructions represents a primary resource to investigate anatomical comparisons and morphometric analysis of cells (Kalisman et al., 2003; Ascoli et al., 2007; Scorcioni et al., 2008; Schmitz et al., 2011; Ramaswamy et al., 2012; Oswald et al., 2013; Muralidhar et al., 2014). It often represents the basis of modeling efforts to study the impact of a cell's morphology on its electrical behavior (Insel et al., 2004; Druckmann et al., 2012; Gidon and Segev, 2012; Bar-Ilan et al., 2013) and on the network it is embedded in (Hill et al., 2012). Many different frameworks, tools and analysis have been developed to contribute to this effort (Schierwagen and Grantyn, 1986; Ascoli et al., 2001, 2007; Ascoli, 2002a,b, 2006; van Pelt and Schierwagen, 2004; Halavi et al., 2008; Scorcioni et al., 2008; Cuntz et al., 2011; Guerra et al., 2011; Schmitz et al., 2011; Hill et al., 2012), such as the Carmen project, framework focusing on neural activity (Jessop et al., 2010), NeuroMorpho.org, repository of digitally reconstructed neurons (Halavi et al., 2008) or the TREES toolbox for morphological modeling (Cuntz et al., 2011).

In spite of over a century of research on cortical circuits, it is still unclear how many classes of cortical neurons exist. Neuronal classification remains a challenging topic since it is unclear how to designate a neuronal cell class and what are the best features to define them by (DeFelipe et al., 2013). Recently, quantitative methods using supervised and unsupervised classifiers have become standard for neuronal classification based on morphological, physiological, or molecular characteristics. They provide quantitative and unbiased identification of distinct neuronal subtypes, when applied to selected datasets (Cauli et al., 1997; Karube et al., 2004;

Ma, 2006; Helmstaedter et al., 2008; Karagiannis et al., 2009; McGarry et al., 2010; DeFelipe et al., 2013).

However, more robust classification methods are needed for increasingly complex and larger datasets. As an example, traditional cluster analysis using Ward's method has been effective, but has drawbacks that need to be overcome using more constrained algorithms. More recent methodologies such as affinity propagation (Santana et al., 2013) outperformed Ward's method but on a limited number of classes with a set of interneurons belonging to four subtypes. One important aspect of classifiers is the assessment of the algorithms, crucial for the robustness (Rosenberg and Hirschberg, 2007), particularly for unsupervised clustering. The limitations of morphological classification (Polavaram et al., 2014) are several: the geometry of individual neurons that varies significantly within the same class, the different techniques used to extract morphologies such as imaging, histology, and reconstruction techniques that impact the measures; but also inter-laboratory variability (Scorcioni et al., 2004). Neuroinformatics tools, computational approaches, and openly available data such as that provided by NeuroMorpho.org enable development and comparison of techniques and accuracy improvement.

The objectives of this paper were to improve our knowledge on automatic classification of neurons and to develop and integrate, based on existing tools, a pipeline that goes from morphological features extraction using l-measure (Scorcioni et al., 2008) to cell classification. The accuracy of machine learning classification algorithms was assessed and compared. Algorithms were trained on 430 digitally reconstructed neurons from NeuroMorpho.org (Ascoli et al., 2007), subjectively classified into layers and/or m-types using young and/or adult development state population of the somatosensory cortex in rats. This study shows the results of applying unsupervised and supervised classification techniques to neuron classification using morphological features as predictors.

## MATERIALS AND METHODS

### Data Sample

The classification algorithms have been trained on 430 digitally reconstructed neurons (**Figures 1** and **2**) classified into a maximum of 22 distinct layers and/or m-types of the somatosensory cortex in rats (**Supplementary Datasheet S1**), obtained from NeuroMorpho.org (Halavi et al., 2008) searched by Species (Rat), Development (Adult and/or Young) and Brain Region (Neocortex, Somatosensory, Layer 2/3, Layer 4, Layer 5, and Layer 6). Established morphological criteria and nomenclature created in the last century were used. In some cases, the m-type names reflect a semantic convergence of multiple given names for the same morphology. We added a prefix to the established names to distinguish the layer of origin (e.g., a Layer 4 Pyramidal Cell is labeled L4_PC). Forty-three morphological features[1] were extracted for each neuron using L-Measure tool (Scorcioni et al., 2008) that provide extraction of quantitative morphological measurements from neuronal reconstructions.

### Data Preprocessing

The datasets were preprocessed (**Supplementary Figure S1**) in order to deal with missing values, often encoded as blanks, NaNs or other placeholders. All the missing values were replaced using the mean value of the processed feature for a given class. Categorical features such as morphology types were encoded transforming each categorical feature with m possible values into m binary features. The algorithms were implemented using different normalization methods depending on the algorithm used. They encompass for scaling feature values to lie between 0 and 1 in order to include robustness to very small standard deviations of features and preserving zero entries in sparse data, normalizing the data using the l2 norm $[-1,1]$ and standardizing the data along any axis.

### Supervised Learning

In supervised learning, the neuron feature measurements (training data) are accompanied by the name of the associated neuron type indicating the class of the observations. The new data are classified based on the training set (Vapnik, 1995). The supervised learning algorithms which have been compared are the following:

– Naive Bayes (Rish, 2001; Russell and Norvig, 2003a,b) with Gaussian Naïve Bayes algorithm (GNB) and Multinomial Naïve Bayes (MNB)
– k-Nearest Neighbors (Cover and Hart, 1967; Hart, 1968; Coomans and Massart, 1982; Altman, 1992; Terrell and Scott, 1992; Wu et al., 2008)
– Radius Nearest Neighbors (Bentley et al., 1977)
– Nearest centroid classifier (NCC) (Tibshirani et al., 2002; Manning, 2008; Sharma and Paliwal, 2010)
– Linear discriminant analysis (LDA) (Fisher, 1936; Friedman, 1989; Martinez and Kak, 2001; Demir and Ozmehmet, 2005; Albanese et al., 2012; Aliyari Ghassabeh et al., 2015)
– Support vector machines (SVM) (Boser et al., 1992; Guyon et al., 1993; Cortes and Vapnik, 1995; Vapnik, 1995; Ferris and Munson, 2002; Meyer et al., 2003; Lee et al., 2004; Duan and Keerthi, 2005) including C-Support Vector Classification (SVC) with linear and radial basis functions (RBF) kernels (SVC-linear and SVC–RBF)
– Stochastic Gradient Descent (SGD) (Ferguson, 1982; Kiwiel, 2001; Machine Learning Summer School and Machine Learning Summer School, 2004)
– Decision Tree (DT) (Kass, 1980; Quinlan, 1983, 1987; Rokach, 2008)
– Random forests classifier[2] (Ho, 1995, 1998; Breiman, 2001)
– Extremely randomized trees[2] (Shi et al., 2005; Geurts et al., 2006, Shi and Horvath, 2006; Prinzie and Van den Poel, 2008)
– Neural Network (McCulloch and Pitts, 1943; Farley and Clark, 1954; Rochester et al., 1956; Fukushima, 1980; Dominic et al., 1991; Hoskins and Himmelblau, 1992) : Multilayer perceptron

---

[1]http://cng.gmu.edu:8080/Lm/help/index.htm

[2]http://scikit-learn.org, 2012

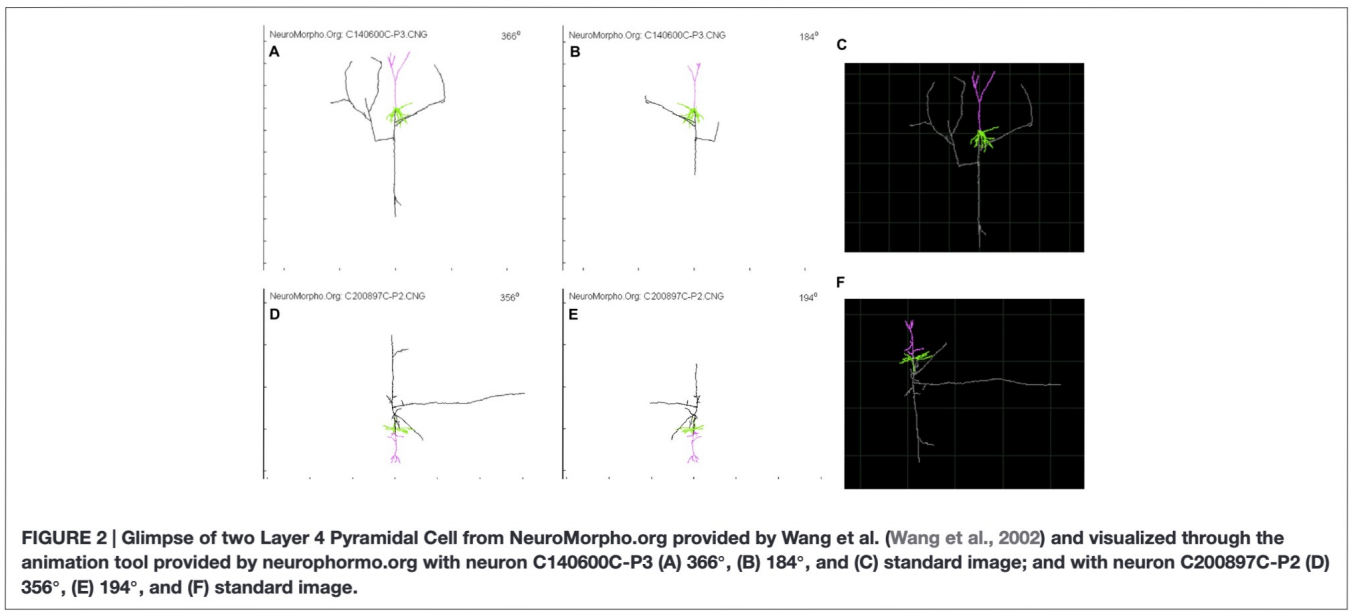

FIGURE 1 | Data Sample of the 430 neurons with Layer 2/3, 4, 5 and 6 as L23, L4, L5 and L6 and m-type Basket, Bipolar, Bitufted, Chandelier, Double Bouquet, Martinotti, Pyramidal, and Stellate cells as BC, BPC, BTC, ChC, DBC, MC, PC, and SC.

FIGURE 2 | Glimpse of two Layer 4 Pyramidal Cell from NeuroMorpho.org provided by Wang et al. (Wang et al., 2002) and visualized through the animation tool provided by neurophormo.org with neuron C140600C-P3 (A) 366°, (B) 184°, and (C) standard image; and with neuron C200897C-P2 (D) 356°, (E) 194°, and (F) standard image.

(MLP) (Auer et al., 2008) and Radial Basis Function network (Park and Sandberg, 1991; Schwenker et al., 2001)
- Classification and Regression Tree (C&R Tree) (Breiman et al., 1984)
- CHi-squared Automatic Interaction Detector (CHAID) (Kass, 1980)
- Exhaustive CHAID (Biggs et al., 1991)
- C5.0 (Patil et al., 2012).

The algorithm for neuron classification using supervised classifier is shown in **Algorithm 1**.

**Algorithm 1**: Neuron supervised classification

1. Normalize each of the neuron feature values
2. Instantiate the estimator
3. Fit the model according to the given training data and parameters
4. Assign to all neurons the class determined by its exemplar
5. Compute the classification accuracy (cf. classification assessment)

The normalization has been chosen regarding the requirements of the classification methods and/or providing the best results.

## Unsupervised Learning

Ten unsupervised (clustering and dimension reduction) algorithms were implemented and assessed. In unsupervised learning classification, the class labels of training data are unknown and the aim is to establish the existence of classes or clusters in the data given a set of measurements. The unsupervised learning algorithms which were compared are the following:

- K-Means (MacQueen, 1967; Lloyd, 1982; Duda, 2001; Kanungo et al., 2002; MacKay, 2003; Jain, 2010; Vattani, 2011; Cordeiro de Amorim and Mirkin, 2012; de Amorim and Hennig, 2015)
- Mini Batch K-Means (Sculley, 2010)
- The K-Means algorithm has been also used on PCA-reduced data (Ding and He, 2004)

- Ward (Ward, 1963; Everitt, 2001; de Amorim, 2015) with and without connectivity constraints
- Mean Shift (Fukunaga and Hostetler, 1975; Cheng, 1995; Comaniciu and Meer, 2002)

The algorithm for the classifiers described above is shown in **Algorithm 2**.

**Algorithm 2**: Unsupervised classification

1. Normalize each of the neuron feature values using the l2 norm [−1,1]
2. Instantiate the estimator
3. Fit the model according to the given training data and parameters
4. Assign to all neurons the corresponding cluster number
5. Algorithm Assessment (cf. classification assessment)

We also used the **Affinity propagation algorithm** (Frey and Dueck, 2007; Vlasblom and Wodak, 2009; Santana et al., 2013) which creates clusters by sending messages between pairs of samples until convergence. Two affinity propagation algorithms have been computed. The first one is based on the Spearman distance, i.e., the Spearman Rank Correlation measures the correlation between two sequences of values. The second one is based on the Euclidian distance. For both, the similarity measure is computed as the opposite of the distance or equality. The best output was kept. The preference value for all points is computed as the median value of the similarity values. Several preference values have been tested including the minimum and mean value. The parameters of the algorithm were chosen in order to provide the closest approximation between the number of clusters from the output and the true classes.

The affinity propagation algorithm is shown in **Algorithm 3**.

**Algorithm 3**: Affinity Propagation

1. Normalize each of the neuron feature values to values [0,1]
2. Similarity values between pairs of morphologies using Spearman/Euclidian distance
3. Calculate the preference values for each morphology
4. Affinity propagation clustering of morphologies
5. Assign to all neurons the corresponding copy number
6. Algorithm Assessment (cf. classification assessment)

For all the clustering algorithms, the exemplar/centroids/cluster number determines the label of all points in the cluster, which is then compared with the true class of the morphology.

**Principal Component Analysis** (PCA; Hotelling, 1933; Abdi and Williams, 2010; Albanese et al., 2012) has been also tested.

## Hardware and Software

The algorithms were implemented in Python 2.7[3] using the Scikit-learn[4] open source python library. Scikit-learn is an open source tool for data mining and data analysis built on NumPy, a package for scientific computing with Python[5], and Scipy, an open source software package for mathematics, science and engineering[6]. Morphological features[7] were extracted for each neuron using L-Measure tool (Scorcioni et al., 2008) allowing extracting quantitative morphological measurements from neuronal reconstructions. The pipeline described above, featuring extraction with l-measure and classification algorithms, was implemented and tested on Power 8 from IBM (S822LC+GPU (8335-GTA), 3.8 GHz, RHEL 7.2). The code of the pipeline is avalaible in GitHub at https://github.com/xaviervasques/Neuron_Morpho_Classification_ML.git.

## Classification Assessment

For supervised algorithm assessment, accuracy statistics have been computed for each algorithm:

$$\text{accuracy} = \frac{\text{corrected labels}}{\text{total samples}}. \quad (1)$$

The accuracy ranges from 0 to 1, with 1 being optimal. In order to measure prediction performance, also known as model validation, we computed a 10 times cross validation test using a randomly chosen subset of 30% of the data set and calculated the mean accuracy. This gives a more accurate indication on how well the model is performing. Supervised algorithms were tested on neurons gathered by layers and m-types in young and adult population, by layers and m-types in young population, by m-types in young and adult population, by m-types in young population, and by layers of Pyramidal Cells in young population. We also performed the tests by varying the percentage of train to test the ratio of samples from 1 to 80%. We provided the respective standard deviations. We included in the python code not only the accuracy code which is shown in this study but also the recall score, the precision score and the F-measure scores. We also built miss-classification matrices for the algorithm providing the best accuracy for each of the categories with the true value and predicted value with the associated percentage of accuracy. In order to assess clustering algorithm, the V-measure score was used (Rosenberg and Hirschberg, 2007). The V-measure is actually equivalent to the normalized mutual information (NMI) normalized by the sum of the label entropies (Vinh et al., 2009; Becker, 2011). It is a conditional entropy-based external cluster measure, providing an elegant solution to many problems that affects previously defined cluster evaluation measures. The V-measure is also based upon two criteria, homogeneity and completeness (Rosenberg and Hirschberg, 2007) since it is computed as the harmonic mean of distinct homogeneity and completeness scores. The V-measure has been chosen as our main index to assess the unsupervised algorithm. However, in order to facilitate the comparison between supervised and unsupervised algorithms, we computed additional types of scores[4] including:

- Homogeneity: each cluster contains only members of a single class.

---

[3]http://www.python.org
[4]http://scikit-learn.org/stable/
[5]http://www.numpy.org
[6]http://scipy.org
[7]http://cng.gmu.edu:8080/Lm/help/index.htm

- Completeness, all members of a given class are assigned to the same cluster.
- Silhouette Coefficient: used when truth labels are not known, which is not our case, and which evaluates the model itself, where a higher Silhouette Coefficient score relates to a model with better defined clusters.
- Adjusted Rand Index: given the knowledge of the ground truth class assignments and our clustering algorithm assignments of the same samples, the **adjusted Rand index** is a function that measures the **similarity** of the two assignments, ignoring permutations and **including chance normalization.**
- Adjusted Mutual Information: given the knowledge of the ground truth class assignments and our clustering algorithm assignments of the same samples, the **Mutual Information** is a function that measures the **agreement** of the two assignments, ignoring permutations. We used **specifically Adjusted Mutual Information.**

The PCA have been assessed using the explained variance ratio.

## RESULTS

### Supervised Algorithms Assessment

The assessment of the supervised algorithms is shown in **Figure 3**. The results showed that LDA is the algorithm providing the best results in all categories when classifying neurons according to layers and m-types with adult and young population (Accuracy score of $0.9 \pm 0.02$ meaning $90\% \pm 2\%$ of well classified neurons), layers and m-type with young population ($0.89 \pm 0.03$), m-types with adult and young population ($0.96 \pm 0.02$), m-types with young population ($0.95 \pm 0.01$), and layers on pyramidal cells with young population ($0.99 \pm 0.01$). Comparing the means between LDA algorithm and all the other algorithms using $t$-test showed a significant difference ($p < 0.001$) and also for C&R Tree algorithm using the Wilcoxon statistical test ($p < 0.005$). We also performed the tests by varying the percentage of train to test the ratio of samples from 1 to 80% showing a relative stability of the LDA algorithm through its accuracy scores and respective standard deviations (**Figure 4**). **Table 1** shows mean precision, recall and F-Scores of the LDA algorithm with their standard deviations for all the categories tested. We also built miss-classification matrices for the LDA algorithm which provided the best accuracy for each of the categories (**Figure 5**).

### Unsupervised Algorithms Assessment

The results of the unsupervised clustering algorithm assessment are shown in **Figure 6**. The algorithms providing slightly better results are the affinity propagation (spearman) and the Ward. The results showed that affinity propagation with Spearman distance is the algorithm providing the best results in neurons classified according to layers and m-types with young and adult population ($V$-measure of 0.44, 36 clusters), layers on pyramidal cells with young population ($V$-measure of 0.385, 22 clusters) and also layers and m-types with young population ($V$-measure of 0.458, 28 clusters). The results showed that Ward algorithms provide the best results on two categories, namely the classification of morphologies according to m-types with young and adult population ($V$-measure of 0.562, 8 clusters), and m-types with young population ($V$-measure 0.503, 8 clusters). Affinity propagation with Euclidean distance has been the second best for all the categories except neurons classified according to Layers and m-types with young and adult population.

## DISCUSSION

Neuron classification remains a challenging topic. However, increasing amounts of morphological, electrophysiological (Sills et al., 2012) and molecular data will help researchers to develop better classifiers by finding clear descriptors throughout a large amount of data and increasing statistical significance. In line with this statement, data sharing and standardized approaches are crucial. Neuroinformatics plays a key role by working on gathering multi-modal data and developing methods and tools allowing users to enhance data analysis and by promoting data sharing and collaboration such as the Human Brain Project (HBP) and the Blue Brain Project (BBP; Markram, 2012, 2013). One of the HBP's objectives, in particular the Neuroinformatics Platform, is to make it easier for scientists to organize and access data such as neuron morphologies, and the knowledge and tools produced by the neuroscience community.

On the other hand, neuron classification is an example of data increase rendering classification efforts harder rather than easier (DeFelipe et al., 2013). Nowadays, different investigators use their own tools and assign different names for classification of the same neuron rendering classification processes difficult, despite existing initiatives for nomenclature consensus (DeFelipe et al., 2013).

Recently, quantitative methods using supervised and unsupervised classifiers have become standard for classification of neurons based on morphological, physiological, or molecular characteristics. Unsupervised classifications using cluster analysis, notably based on morphological features, provide quantitative and unbiased identification of distinct neuronal subtypes when applied to selected datasets. Recently, Santana et al. (2013) explored the use of an affinity propagation algorithm applied to 109 morphologies of interneurons belonging to four subtypes of neocortical GABAergic neurons (31 BC, 23 ChC, 33 MC, and 22 non-MC) in a blind and non-supervised manner. The results showed an accuracy score of 0.7374 for the affinity propagation and 0.5859 for Ward's method. McGarry et al. (2010) classified somatostatin-positive neocortical interneurons into three interneuron subtypes using 59 GFP-positive interneurons from mouse. They used unsupervised classification methods with PCA and K-means clustering assessed by the silhouette analysis measures of quality. Tsiola et al. (2003) used PCA and cluster analysis using Euclidean distance by Ward's method on 158 neurons of Layer 5 neurons from Mouse Primary Visual Cortex. Muralidhar et al. (2014) and Markram et al. (2015) applied classification methodologies on Layer 1 DAC (16), HAC(19), SAC (14), LAC (11), and NGC (17 NGC-DA and 16 NGC-SA) using objective (PCA and LDA), supervised and unsupervised methodologies on the developing rat's somatosensory cortex using multi-neuron patch-clamp and 3D morphology reconstructions. The study suggested that objective unsupervised classification (using PCA) of neuronal morphologies is currently not possible based on any single independent feature of neuronal morphology of layer 1. The LDA is a supervised method that performs better to classify the L1 morphologies where cell classes were distinctly separated.

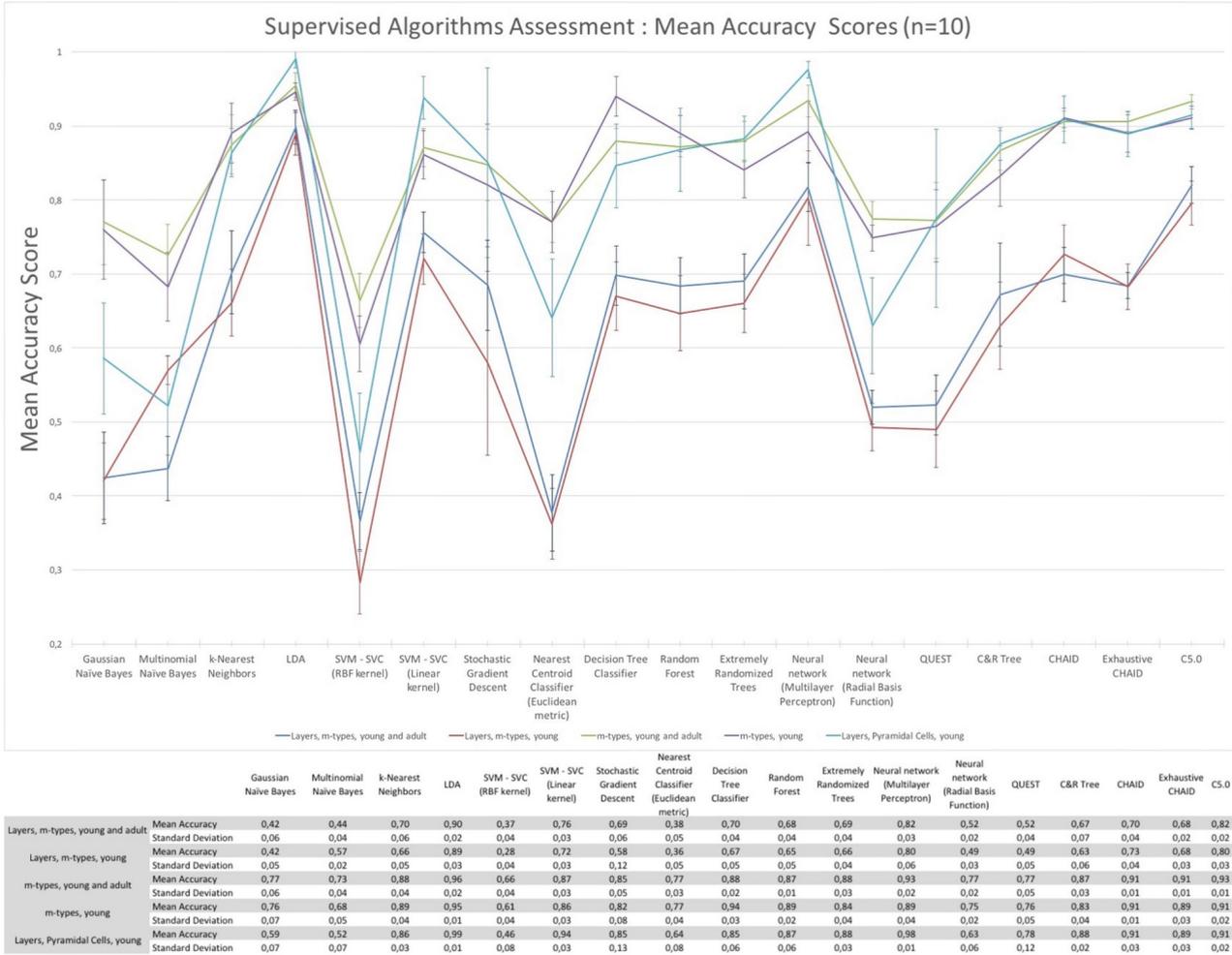

FIGURE 3 | **The mean accuracy scores with their respective standard deviation of the supervised algorithms.** The mean accuracy scores have been computed 10 times using a randomly chosen 30% data subset to classify morphologies according to layers and m-types, m-types, and layers only.

FIGURE 4 | **Tests varying the percentage of train to test the ratio of samples from 1 to 80% showing a relative stability of the linear discriminant analysis (LDA) algorithm.** The figure shows the mean accuracy scores with their respective standard deviation for each of the category tested.

Additional algorithms can be included in the pipeline presented in our study such as nearest neighbor methods, feature decisions based methods, or those using metric learning that might be more robust for the given problem in low numbers of training samples.

The algorithms in the literature are often applied to small amounts of data, few classes and are often specific to a layer. In our study, we assess and compare accuracy of different classification algorithms and classified neurons according to layer and/or m-type with young and/or adult development state in an important training sample size of neurons. We confirm that supervised with LDA is an excellent classifier showing that quality data are mandatory to predict the class of the morphology to be classified. Subjective classification by a neuroscientist is crucial to improve models on curated data. The limitations of morphological classification (Polavaram et al., 2014) are well known, such as the morphometric differences between laboratories, making the classification harder. Further challenges are the geometry of individual neurons which varies significantly within the same class, the different techniques used to extract morphologies (such as imaging, histology, and reconstruction techniques) which impact the measures, and finally, inter-laboratory variability (Scorcioni et al., 2004). Neuroinformatic tools, computational approaches and their open availability, and data such as NeuroMorpho.org make it possible to develop and compare techniques and improve accuracy.

TABLE 1 | Mean precision, recall and *F*-scores of the linear discriminant analysis (LDA) algorithm with their respective standard deviations for all the categories tested.

| Mean scores | Precision | Recall | *F*-score |
|---|---|---|---|
| Layers, m-types: young and adult | 0.9 ± 0.02 | 0.91 ± 0.015 | 0.9 ± 0.022 |
| Layers, m-types: young | 0.87 ± 0.03 | 0.88 ± 0.03 | 0.86 ± 0.03 |
| m-types: young and adult | 0.95 ± 0.02 | 0.94 ± 0.03 | 0.94 ± 0.03 |
| m-types: young | 0.94 ± 0.02 | 0.94 ± 0.01 | 0.94 ± 0.01 |
| Layer, pyramidal cells: young | 0.98 ± 0.01 | 0.98 ± 0.01 | 0.98 ± 0.01 |

All unsupervised algorithms applied to a larger number of neurons showed, as expected, lower results. One limitation of these algorithms is that results are strongly linked to parameters of the algorithm that are very sensible. The current form of affinity propagation suffers from a number of drawbacks such as robustness limitations or cluster-shape regularity (Leone et al., 2007), leading to suboptimal performance. PCA gives poor results confirming the conclusion of Muralidhar et al. (2014)

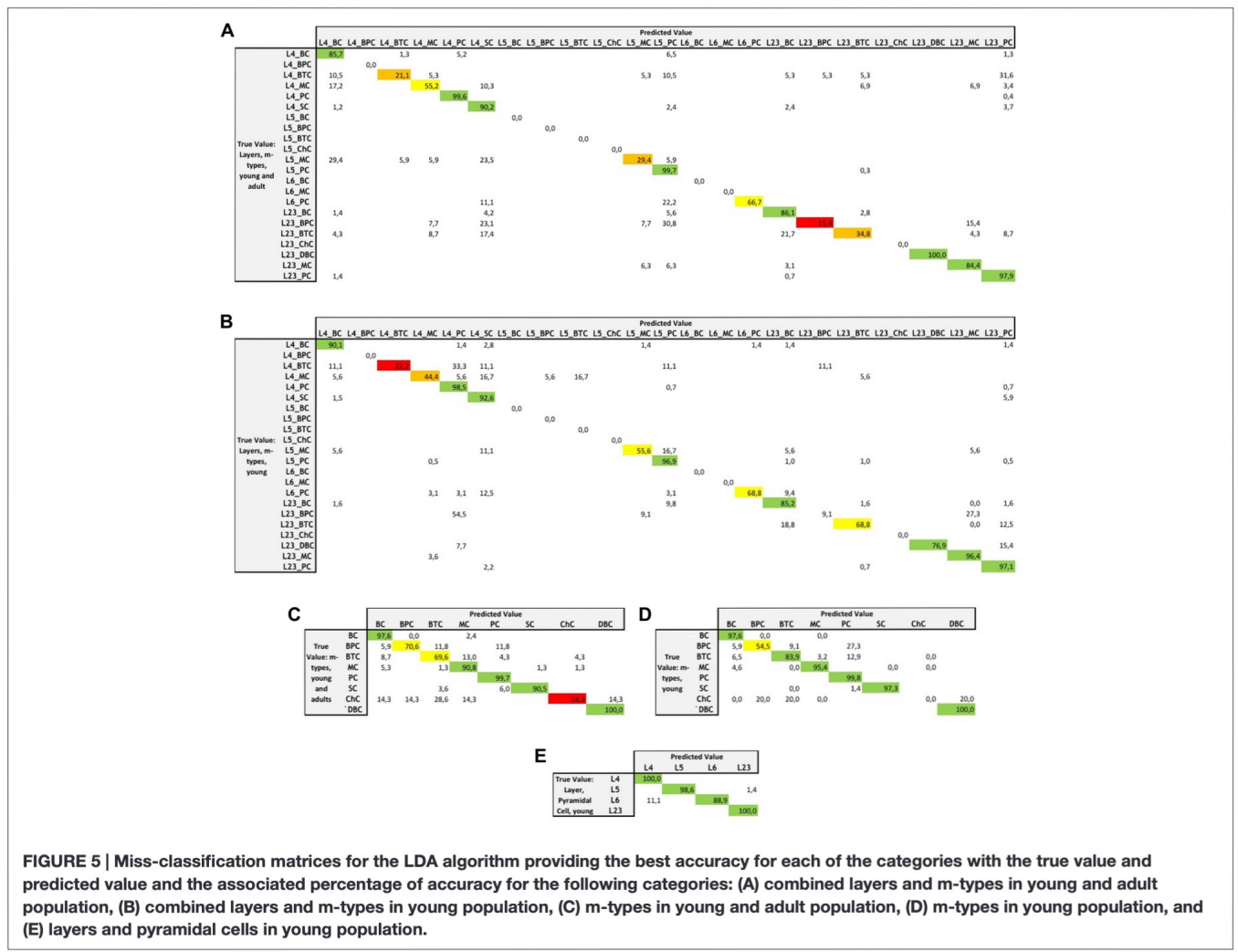

FIGURE 5 | Miss-classification matrices for the LDA algorithm providing the best accuracy for each of the categories with the true value and predicted value and the associated percentage of accuracy for the following categories: (A) combined layers and m-types in young and adult population, (B) combined layers and m-types in young population, (C) m-types in young and adult population, (D) m-types in young population, and (E) layers and pyramidal cells in young population.

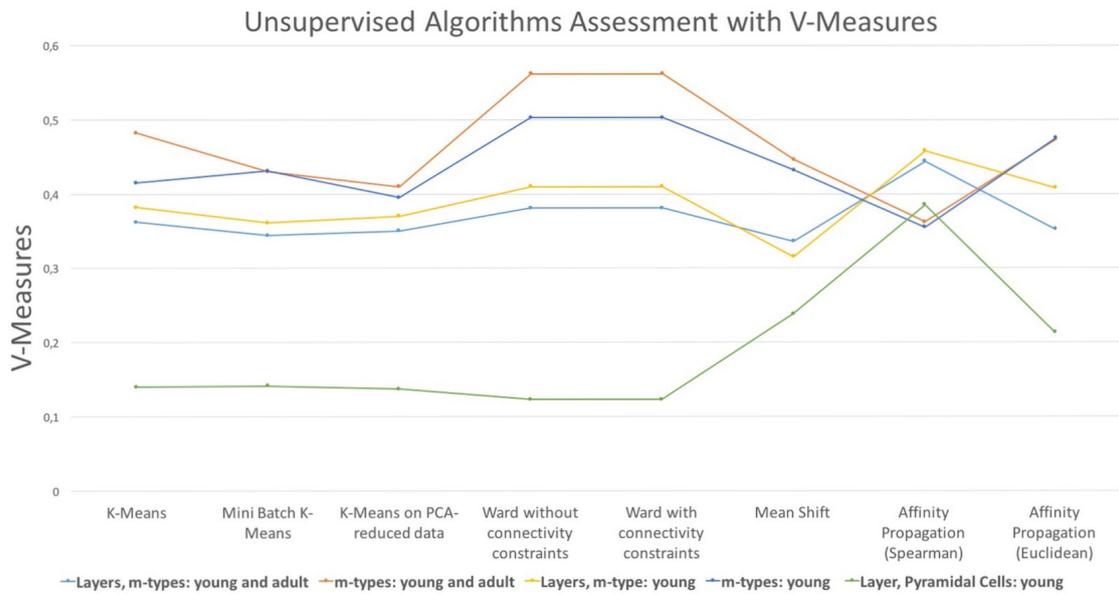

**FIGURE 6 | V-measures comparison of the unsupervised clustering algorithms classifying morphologies according to layer and m-type, m-type and layer in young an/or adult population.** The figure shows also homogeneity scores, completeness scores, adjusted rand index, adjusted mutual information, and silhouette coefficient.

that PCA could not generate any meaningful clusters with a poor explained variance in the first principal component. Objective unsupervised classification of neuronal morphologies is currently tricky even using similar unsupervised methods.

Supervised classification seems the best way to classify neurons according to previous subjective classification of neurons. Data curation is critically important. Brain modeling efforts will require huge volumes of standardized, curated data on the brain's different levels of organization as well as standardized tools for handling them.

One important aspect of classifiers is the evaluation of the algorithms to assess the robustness of classification methods (Rosenberg and Hirschberg, 2007), especially for unsupervised clustering. The silhouette analysis measure of quality (Rousseeuw, 1987) is one of them. More recently, the use of $V$-measure, an external entropy based cluster evaluation measure, provides an elegant solution to many problems that affect previously defined cluster evaluation measures (Rosenberg and Hirschberg, 2007). Features selection would show important features to classify morphologies. Taking into account only the significant features to classify morphologies will increase the accuracy of classification algorithms. Features selection is a compromise between running time by selecting the minimum sized subset of features and classification accuracy (Tang et al., 2014). The combination of different types of descriptors is crucial and can serve to both sharpen and validate the distinction between different neurons (Helmstaedter et al., 2008, 2009; Druckmann et al., 2012). We propose that the appropriate strategy is to first consider each type of descriptor separately, and then to combine them, one by one, as the data set increases. Independently analyzing different types of descriptors allows their statistical power to be clearly examined. A next step of this work will be to combine morphology data with other types of data such as electrophysiology and select the significant features which help improve classification algorithms with curated data.

The understanding of any neural circuit requires the identification and characterization of all its components (Tsiola et al., 2003). Neuronal complexity makes their study challenging as illustrated by their classification and nomenclature (Bota and Swanson, 2007; DeFelipe et al., 2013), still subject to intense debate. While this study deals with morphologies obtainable in optic microscope it would be useful to know how the electron microscopic features can also be exploited. Integrating and analyzing key datasets, models and insights into the fabric of current knowledge will help the neuroscience community to face these challenges (Markram, 2012, 2013; Kandel et al., 2013). Gathering multimodal data is a way of improving statistical models and can provide useful insights by comparing data on a large scale, including cross-species comparisons and cross datasets from different layers of the brain including electrophysiology (Menendez de la Prida et al., 2003; Czanner et al., 2008), transcriptome, protein, chromosome, or synaptic (Kalisman et al., 2003; Hill et al., 2012; Wichterle et al., 2013).

## AUTHOR CONTRIBUTIONS

XV is the main contributor of the article from the development of the code to results, study of the results, and writing article. LC supervised and wrote the article. LV and GV did code optimization.

## ACKNOWLEDGMENT

We acknowledge Alliance France Dystonie and IBM France (Corporate Citizenship and Corporate Affairs) for their support.

## SUPPLEMENTARY MATERIAL

The Supplementary Material for this article can be found online at: http://journal.frontiersin.org/article/10.3389/fnana.2016.00102/full#supplementary-material

**FIGURE S1** | Block Diagram of the python pipeline.

**DATASHEET S1** | File containing all the name of the neurons (Neuron_ID.xlsx) from which the particular neurons can be identified and inspected.